\documentclass[a4paper,fleqn]{cas-sc}

\usepackage[authoryear]{natbib}
\usepackage{amsmath,amsfonts}
\usepackage{array}
\usepackage{caption}
\usepackage{float}
\usepackage{graphicx}
\usepackage{booktabs}
\usepackage{xcolor}
\usepackage{soul}
\usepackage{tabularx}
\usepackage{amssymb}
\usepackage{xurl}
\usepackage{placeins}
\usepackage{microtype}

\def\tsc#1{\csdef{#1}{\textsc{\lowercase{#1}}\xspace}}
\tsc{WGM}
\tsc{QE}
\tsc{EP}
\tsc{PMS}
\tsc{BEC}
\tsc{DE}

\begin{document}
\let\WriteBookmarks\relax
\renewcommand{\topfraction}{0.95}
\renewcommand{\bottomfraction}{0.9}
\renewcommand{\textfraction}{0.03}
\renewcommand{\floatpagefraction}{0.75}
\setcounter{topnumber}{3}
\setcounter{bottomnumber}{2}
\setcounter{totalnumber}{6}
\setlength{\textfloatsep}{12pt plus 3pt minus 2pt}
\setlength{\floatsep}{10pt plus 2pt minus 2pt}
\setlength{\intextsep}{10pt plus 2pt minus 2pt}
\setlength{\emergencystretch}{1em}
\clubpenalty=10000
\widowpenalty=10000
\raggedbottom
\makeatletter
\setlength{\@fptop}{0pt}
\setlength{\@fpsep}{14pt plus 2pt minus 2pt}
\setlength{\@fpbot}{0pt plus 1fil}
\makeatother

\shorttitle{Decommissioning Rural Roadside HAR and DMS}

\shortauthors{Chen et~al.}

\title[mode=title]{An Evidence-Based Decision Framework for Retiring Roadside Traveler Information Infrastructure in the Mobile Era}

\tnotemark[1]

\tnotetext[1]{This work was supported by the Maryland DOT State Highway Administration. (Corresponding author: Xianfeng Terry Yang.)}

\author[1]{Dianwei Chen}[orcid=0009-0005-8227-317X]

\ead{dwchen98@umd.edu}

\author[1]{Yuan-Zheng Lei}
\ead{yzlei@umd.edu}
\author[1]{Zheyu Li}
\ead{zli9226@umd.edu}
\author[1]{Xianfeng Terry Yang}[orcid=0000-0002-9416-6882]
\cormark[1]
\ead{xtyang@umd.edu}

\affiliation[1]{
    organization={Department of Civil and Environmental Engineering, University of Maryland},
    addressline={College Park},
    city={College Park},
    state={MD},
    postcode={20740},
    country={USA}
}

\cortext[cor1]{Corresponding author}

\begin{abstract}
As mobile traveler-information services expand, transportation agencies must determine whether aging roadside infrastructure can be retired while preserving public access and operational resilience. This study develops an evidence-based decision framework for screening highway advisory radio (HAR) and dynamic message sign (DMS) assets for retirement. The framework integrates carrier-specific connectivity measurements, asset inventories, and documented mobile-message receipts, while separating technical evidence, institutional decision rules, and requirements for subsequent agency review. A Maryland case study combines three-carrier drive tests across four corridors with an inventory of nine HAR sites and 234 DMS. Alternative screening scenarios examine how evidence availability, radio references, and institutional assumptions affect review categories. All 105 DMS with complete local radio observations pass the baseline connectivity screen, leaving inventory rules to differentiate their action-review categories. In analyst-defined scenarios, removing ownership-based retention increases decommissioning referrals from three to 21, whereas extending beacon retention to urban assets reduces them to zero. These results demonstrate that retirement screening depends substantially on institutional assumptions, even when the underlying radio evidence remains unchanged. Explicit treatment of missing and conditional evidence also identifies assets requiring further assessment. Because historical radio measurements are spatially associated with later message tests, the resulting references remain exploratory and do not establish delivery reliability. The framework provides an auditable basis for prioritizing asset reviews and documenting decision assumptions; final retirement decisions require asset-specific assessments of service equivalence, audience reach, resilience, and lifecycle costs.

\end{abstract}


\maketitle

\section{Introduction}
\label{sec:intro}

Transportation agencies increasingly face decisions about the future of aging roadside traveler-information infrastructure as mobile information services expand. Highway advisory radio (HAR) and dynamic message signs (DMS) have long provided direct channels for communicating traffic conditions, incidents, and other agency messages to travelers. Mobile navigation, 511 services, agency alerts, and third-party platforms now provide many of the same types of information, while legacy roadside equipment continues to require maintenance and replacement investment \citep{raman2006communicating, dudek2004changeable, mckay2022applying}. This transition raises a practical policy question: under what conditions should agencies retain, modernize, convert, or retire existing roadside assets? Answering this question requires an assessment of whether alternative communication channels can preserve the public service provided by each asset.

The service implications of retirement are particularly important on rural roads, where cellular connectivity varies by carrier and terrain \citep{adarsh2021coverage, mangla2021tale}. HAR broadcasts information over a corridor, whereas a DMS presents an agency message at a specific roadway location. Mobile alerts can provide information farther upstream and tailor content to individual travelers, but their effectiveness depends on network availability, device receipt, and user attention \citep{bean2015study}. Roadside channels also provide access to agency information without requiring an application, data plan, or subscription to a particular carrier. Retirement decisions therefore involve questions of audience reach, timely warning, and operational resilience. For example, a message received after the final feasible diversion point may have little practical value, even when technical delivery succeeds. Agencies must consider these service requirements alongside equipment condition and potential maintenance savings.

Existing research and guidance provide useful foundations for addressing this decision. The FHWA and AASHTO asset-management guidance reviewed in this study emphasizes equipment condition, performance, risk, and lifecycle planning \citep{mckay2022applying, american2011transportation}. Studies of DMS, mobile alerts, and crowdsourced traveler information examine the operation and performance of individual communication channels \citep{wardman1997driver, chatterjee2002driver, amin2018evaluating}. However, these sources do not provide an integrated procedure for connecting carrier-specific field measurements and mobile-message observations to the review of individual roadside assets. In particular, agencies need a transparent way to distinguish measured weaknesses from missing evidence, identify the assumptions underlying screening categories, and determine which service requirements remain unresolved before an asset can advance toward retirement.

This study addresses the following research question: what evidence should transportation agencies require before advancing roadside traveler-information assets to retirement review, and how do alternative evidence requirements and institutional rules affect screening outcomes? We develop an evidence-based decision framework that integrates connectivity measurements, asset inventories, and documented mobile-message receipts. The framework evaluates connectivity at both corridor and asset scales, preserves carrier-specific results, and distinguishes unavailable evidence from observations that fall below the selected screening references. It also records technical evidence separately from inventory-based review categories and subsequent agency decisions. This structure allows agencies to examine how a screening outcome depends on local measurements, the selected evidence requirements, and assumptions about asset ownership and roadside warning functions.

The framework is applied to a Maryland state highway inventory containing 9 HAR sites and 234 DMS, using three-carrier drive tests across four corridors and follow-up SMS and WhatsApp receipt observations. Historical radio measurements are spatially associated with later message-test locations to construct exploratory screening references; this association does not establish radio conditions at the time of receipt. The analysis compares inventory-only, connectivity-only, baseline, and evidence-gated screening scenarios and examines sensitivity to radio references, evaluation footprints, carrier requirements, and institutional rules. Asset-level audits then identify the evidence supporting the baseline decommissioning referrals and the information still required for agency review.

This study makes three contributions:
\begin{enumerate}
    \item It develops an auditable decision framework that separates connectivity evidence, preliminary asset-review categories, and agency judgment. The framework explicitly distinguishes missing, weak, and conditional evidence and records the service requirements that must be addressed before an implementation decision.
    
    \item It quantifies how local evidence requirements and screening assumptions affect the assets eligible for further review. The Maryland application combines carrier-specific corridor and asset assessments with sensitivity analyses to identify where screening results remain stable and where additional evidence is needed.
    
    \item It evaluates how institutional rules concerning ownership and roadside warning features influence screening outcomes. Comparisons across alternative rules and audits of individual referrals clarify which outcomes depend on radio evidence and which depend on the assumptions governing asset review.
\end{enumerate}

The framework provides a basis for prioritizing further assessment and documenting the reasoning behind asset decisions. Its transferable contribution is the explicit connection between available evidence, screening assumptions, and unresolved service requirements. Agencies can update these records as new measurements and operational information become available, while retaining a traceable basis for subsequent reviews of service equivalence, audience reach, resilience, and lifecycle costs.

\section{Literature review}\label{sec:literature}

\subsection{Roadside and mobile traveler-information channels}

HAR and DMS deliver agency messages without requiring a personal device, subscription, or installed application. HAR provides corridor broadcasts, while DMS support incident, work-zone, route-choice, and travel-time messaging \citep{raman2006communicating, dudek2004changeable}. Mobile navigation, 511 systems, agency applications, and crowdsourced traffic feeds add channels that can deliver information before a traveler reaches a roadside asset \citep{raman2006communicating, amin2018evaluating}. The channels are therefore complementary, but they do not provide the same access or spatial reach.

Technical delivery alone does not establish that a warning is usable. Research on DMS emphasizes brevity, credibility, location, legibility, and the action requested of the driver \citep{dudek2004changeable, weaver2024systematic, kondyli2026creating}. Studies of mobile public warnings similarly examine message content, source authority, and the recipient's ability to understand and act on an alert \citep{bean2015study, cain2024wireless, olson2024decade}. A substitution analysis must therefore consider whether the message arrives through an authoritative channel and before the last point at which the traveler can take the intended action.

Evidence from DMS studies also depends on the outcome being examined. DMS messages have been associated with route-choice responses in urban networks and speed changes on a rural interstate \citep{wardman1997driver, chatterjee2002driver, sui2014impact}. These behavioral findings do not show that a mobile channel can replace a particular sign. Mobile receipt also depends on radio availability, device and platform settings, message design, and enrollment in the relevant service \citep{bean2015study, cain2024wireless, olson2024decade}. The replacement question must consequently be evaluated for the function and decision point of the individual asset rather than inferred from general technology adoption.

The required evidence also changes with message type. Routine travel times, planned work-zone information, sudden closures, weather warnings, and evacuation instructions differ in urgency, audience, and need for backup. A channel that is adequate for a recurring low-consequence message may not be adequate for a safety-critical use. Asset review should therefore begin with the service historically provided at that location, including infrequent high-consequence messages, and then test whether the proposed channel combination preserves that service.

\subsection{Rural connectivity, access, and redundancy}

Rural mobile coverage varies by carrier, terrain, and settlement pattern, with documented gaps in sparsely served areas \citep{adarsh2021coverage, mangla2021tale}. A corridor mean can conceal both single-carrier gaps and locations where all measured carriers are weak. Likewise, a best-carrier map shows the strongest available network at a location but not the service available to a traveler subscribed to another carrier. Carrier-specific results and jointly matched locations are therefore needed alongside any best-carrier summary.

Radio availability is only one part of mobile access. Digital-divide research distinguishes access to a device or connection from the skills and conditions needed for effective use \citep{mossberger2003virtual, van2020digital}. Smartphone ownership also differs across demographic groups \citep{perrin2021mobile}. Replacing a roadside channel with mobile delivery may therefore shift requirements to travelers: they must have a compatible device and service, enable the relevant alert path, and recognize and act on the message while driving. Statewide adoption rates cannot establish these conditions for the audience passing a particular rural asset.

Emergency-communication research emphasizes the use of multiple pathways to reach affected populations \citep{raman2006communicating, edwards2022best}. However, two outputs are not independent merely because they appear on different devices. They may share event detection, operator entry, power, backhaul, cloud services, or a commercial carrier. Roadside equipment has its own common failure modes, including local power, communication, and controller failures. A redundancy review must therefore document shared dependencies and credible backup paths instead of counting nominal channels \citep{leveson2012engineering}.

The relevant audience may include residents, commuters, freight operators, seasonal visitors, and travelers unfamiliar with an agency application. Their carriers, devices, language needs, and alert settings may differ. For each asset, the agency must determine whom the existing service reaches, which users the proposed replacement may exclude, and whether the message can still be received in time to act.

\subsection{Asset-management gap and study position}

Transportation asset-management guidance organizes ITS decisions around inventory, condition, performance, risk, lifecycle planning, and replacement needs \citep{mckay2022applying, american2011transportation}. Multi-criteria transport research provides methods for structuring decisions with competing technical, economic, environmental, and social criteria \citep{broniewicz2025application}. These bodies of work support disciplined decision making, but they do not by themselves specify how carrier-specific field measurements and observed mobile receipt should enter the review of an individual HAR or DMS.

The distinction matters because equipment condition and service substitutability answer different questions. Poor condition may justify investment, but it does not determine whether the asset should be replaced in kind, modernized, converted to another roadside treatment, or removed. Conversely, strong cellular measurements do not show that a mobile alternative reaches the intended audience, arrives before the decision point, or preserves necessary backup. Lifecycle costs become comparable only after the agency identifies alternatives that meet these service requirements.

Missing measurements must also remain separate from observed weak service. An asset outside the measured area has not been evaluated; an asset with low local measurements provides evidence of a weak mobile alternative under the campaign conditions. Replacing absent local evidence with a corridor or statewide average would obscure the need for additional measurement. The proposed procedure therefore records unmeasured assets as unevaluated and reports sensitivity to the historical spatially associated radio reference and asset footprint.

The study connects these requirements in an asset-level workflow. It combines a geocoded inventory, carrier-specific corridor and footprint measurements, historical radio values spatially associated with later message-test locations, an explicit DMS review tree, and three aspects of agency review. The inventory variables used by the tree are visible proxies rather than direct measures of safety benefit. Evidence states and review categories remain separate from agency decisions so that message history, audience access, operational knowledge, redundancy, and implementation constraints can be reviewed without overwriting the measured evidence.

\FloatBarrier

\section{Method}
\label{sec:method}

\subsection{Screening workflow and data}\label{sec:method:overview}

\begin{figure}[pos=!htbp]
    \centering
    \includegraphics[width=0.96\linewidth]{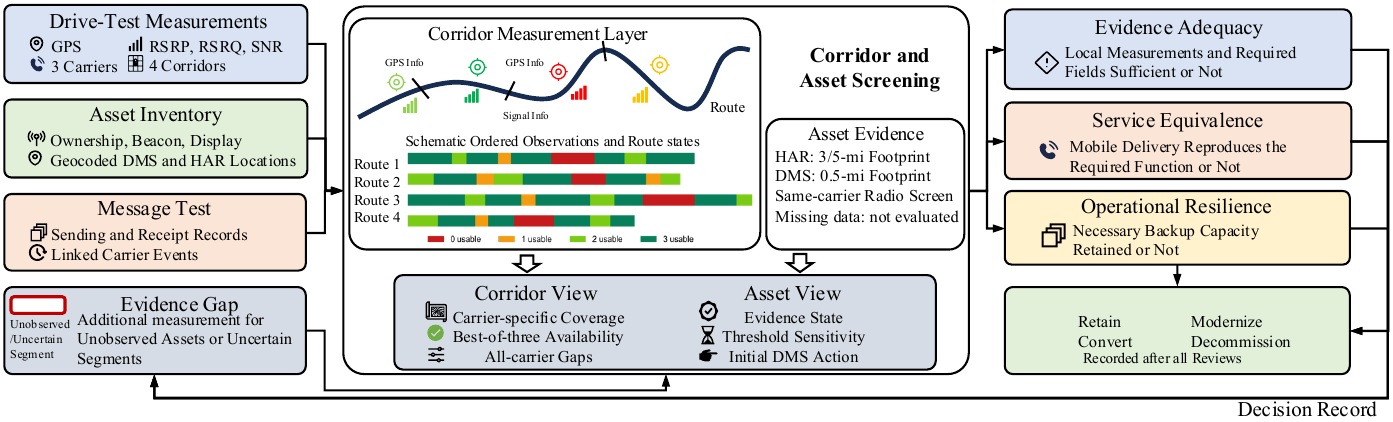}
    \caption{Evidence used in the asset review. The 2025 drive-test layer supplies corridor measurements, asset-footprint measurements, and the historical radio values spatially associated with 2026 message-test locations. Receipt archives identify the sending times retained for the historical association. Screening then informs three agency review questions before a final action is recorded.}
    \label{fig:framework}
\end{figure}

Figure~\ref{fig:framework} summarizes the analysis. Corridor measurements describe carrier-specific signal conditions and identify matched locations where all three measured networks record low connectivity scores. Locations with documented message receipt in 2026 are spatially associated with the 2025 drive-test layer to construct historical radio references for the asset screen. The DMS decision tree combines the baseline screen with inventory fields to assign a preliminary review category. Agency staff separately review evidence adequacy, service equivalence, and implementation resilience before approving an asset-specific action.

Three data sources are used. The geocoded inventory contains HAR location and communication status and DMS location, ownership, beacon availability, and display size. The 2025 drive-test layer contains GPS-aligned RSRP, RSRQ, and SNR records at one-second intervals for AT\&T, T-Mobile, and Verizon. The main corridor and asset analyses retain only records containing all three radio fields. The 2026 message tests provide 156 georeferenced sending locations, including 47 on the Eastern Shore. Archived receipt records can be linked to 109 sending times; 101 of those times have at least one available radio component from the historical spatial lookup. The lookup values and the asset-footprint measurements originate from the same 2025 drive-test layer, so the resulting reference comparison is not an independent validation of message delivery.

The calculations produce results at corridor and asset scales. Corridor results compare carrier-specific signal conditions on the four driven routes. HAR results use measurements within a 3-mi radius as the baseline and a 5-mi radius for sensitivity analysis. For each DMS, the evaluation footprint is a 0.5-mi (804.672-m) radius centered on its inventory coordinate. A DMS is measured in the main analysis when at least one retained complete radio triplet lies inside that circle; the route label used for reporting is the most frequent route among retained observations within 0.5~mi, or within 1~mi when the 0.5-mi circle is empty. An asset without a retained triplet remains unevaluated. This state can reflect either a lack of local sampling or incomplete radio fields, which require different follow-up actions. Section~\ref{sec:method:scoring} describes the location matching used for direct cross-carrier comparisons.

\subsection{Connectivity score and corridor exposure}\label{sec:method:scoring}

Three calculations use different spatial units. Carrier-specific corridor statistics first describe the signal conditions recorded along each route. Direct cross-carrier comparison then uses each retained T-Mobile observation as a reference point. Within the same route, the calculation selects the nearest retained AT\&T and Verizon observations within 5~m. A reference point is excluded when either carrier has no observation within that distance. The matching is spatial rather than temporal and does not require a one-to-one assignment, so one AT\&T or Verizon source row can be selected for more than one nearby T-Mobile observation.

Each complete tri-carrier match is assigned to a route-specific $0.0025^{\circ}$ grid cell using the T-Mobile coordinates. At Maryland latitudes, a cell is approximately 220~m east--west by 280~m north--south. Within each cell, repeated uses of the same source row are removed by observation identifier, and the connectivity scores of the remaining unique observations are averaged separately for each carrier. The 5-m match places the three carrier observations at comparable locations, while the grid aggregation limits the influence of repeated one-second samples. This procedure produces 3{,}723 cells with all three carriers represented.

The asset calculation uses the retained complete drive-test observations rather than the matched cells. Lower-tail RSRP, RSRQ, and SNR values inside each HAR or DMS footprint are compared with historical carrier-specific references associated with receipt locations. Corridor and asset results remain separate because a pass near an asset does not imply a pass throughout the corridor.

Drive-test observations represent conditions during the campaign, not all possible network states. All three carrier devices were mounted in the same vehicle and driven along the same routes, reducing differences in platform position and sampled location. The results remain specific to the campaign dates, network configuration, traffic, foliage, and weather. Complete triplets that produce a zero connectivity score remain in the main analysis. The cleaned route files do not contain a separate service-status field that would establish whether every explicit no-service event has been retained. Excluding incomplete triplets can remove weak observations and alter the sampled locations; a separate score-only sensitivity analysis examines records whose available fields already determine a zero score.

The drive-test dataset records three complementary radio metrics. RSRP measures received signal strength, RSRQ describes signal quality after accounting for interference and network loading, and SNR compares the received signal with background noise. Higher values indicate better radio conditions for all three metrics. Table~\ref{tab:radio_metrics} lists the definition and interpretation of each metric.

\begin{table}[pos=!htbp]
\centering
\caption{Radio metrics used in the connectivity analysis.}
\label{tab:radio_metrics}
\small
\begin{tabularx}{\linewidth}{
    @{}
    l
    >{\hsize=0.75\hsize\raggedright\arraybackslash}X
    >{\hsize=1.25\hsize\raggedright\arraybackslash}X
    @{}
}
\toprule
Metric & Full name & Interpretation \\
\midrule
RSRP &
Reference Signal Received Power &
Received power of the cellular reference signal; primarily represents signal strength. \\

RSRQ &
Reference Signal Received Quality &
Signal quality relative to total received power; reflects interference and network loading. \\

SNR &
Signal-to-Noise Ratio &
Strength of the received signal relative to background noise. \\
\bottomrule
\end{tabularx}
\end{table}

For a measurement $(P,Q,S)=(\mathrm{RSRP},\mathrm{RSRQ},\mathrm{SNR})$, the MDSHA mobile-coverage map converts the three metrics to normalized component scores:
\begin{equation}\label{eq:components}
s_P=\mathrm{clip}_{[0,1]}\!\left(\frac{P+120}{35}\right),\quad
s_Q=\mathrm{clip}_{[0,1]}\!\left(\frac{Q+20}{10}\right),\quad
s_S=\mathrm{clip}_{[0,1]}\!\left(\frac{S}{20}\right),
\end{equation}
The map then combines the components as
\begin{equation}\label{eq:oe}
O(P,Q,S)=100s_P^{0.5}s_Q^{0.2}s_S^{0.3}.
\end{equation}
The map sets $O$ to zero when $P\le-120$~dBm or $S\le0$~dB and bounds the score to $[0,100]$. For each corridor and carrier, the distribution of $O$ and the share of observations below 10 describe overall signal conditions. The main matched-cell comparison uses $O\ge30$, and the threshold sensitivity analysis repeats the calculation from 0 to 50. These values are diagnostic score settings, not validated delivery classes. The asset screen does not use them; it compares raw RSRP, RSRQ, and SNR values with the historical references associated with receipt locations.

A separate score-only sensitivity analysis adds incomplete records when an available RSRP value is at or below $-120$~dBm or an available SNR value is at or below 0~dB. Either condition determines $O=0$ under Eq.~\eqref{eq:oe} without assigning a value to any missing radio component. This adds 3{,}287 records to the corridor-score analysis: 3{,}115 for AT\&T and 172 for T-Mobile. Carrier statistics and the same 5-m matching, within-cell deduplication, and cell-mean calculations are repeated on this expanded score sample. The sensitivity analysis changes neither the raw radio fields nor the complete-triplet sample used for the main asset tests; it is not a reconstruction of all no-service events.

The exponents are inherited from the agency map rather than estimated from the delivery campaigns. RSRP receives the largest weight, while the multiplicative form prevents a strong value on one metric from fully offsetting a near-zero value on another.

For every corridor--carrier pair, the reported statistics include the observation count, mean, median, interquartile range, share with $O<10$, and share with $O\ge70$. The two tail shares distinguish carriers with similar means but different frequencies of poor and strong signal. The primary direct cross-carrier calculations use only the 3{,}723 grid cells constructed from complete tri-carrier matches.

For a retained grid cell $\ell$, let $\overline{O}_{\ell}^{c}$ denote the mean connectivity score for carrier $c$. The best-carrier statistic is
\begin{equation}\label{eq:bestcarrier}
O_\ell^{\star}=\max_{c\in\mathcal C}\overline{O}_\ell^c.
\end{equation}
$O_\ell^{\star}$ is an infrastructure upper bound because it records the strongest measured carrier at location $\ell$; it does not indicate that every traveler can use that carrier. Carrier-specific results are therefore retained. The main comparison uses a score threshold of 30 and reports a \emph{tri-carrier low-score cell} when all three carrier scores fall below it. Because this value is a screening setting rather than a delivery threshold, the analysis is repeated over thresholds from 0 to 50.

The asset analysis compares signal conditions near each HAR or DMS with historical carrier-specific references spatially associated with documented receipt locations. The corridor analysis separately identifies matched locations where all three carrier scores fall below the selected diagnostic setting. An asset may therefore pass its local screen even when poor signal remains elsewhere on the same corridor.

\subsection{Historical radio references associated with receipt locations}\label{sec:method:floor}

The radio fields in the message-test workbooks were obtained by spatial lookup in the 2025 drive-test layer, not by contemporaneous radio measurement during message receipt in 2026. For each carrier, the lookup associates the test GPS position with a nearest eligible historical route record within 200~m; the workbooks retain the assigned route and matching distance. These values describe earlier radio conditions near the test location. The April workbook contains 35 sending times with six archived receipt fields covering three carriers and two channels. For May, sending times normalized to the minute link 74 May~19 and May~21 schedule rows to receipt records for all three carriers. The 47 May~23 locations have no corresponding receipt records in the local archive and are excluded from the historical reference sample.

The receipt-linked records represent 109 sending times and 432 documented successful carrier--channel deliveries. Eight April sending times have no available historical radio component for any carrier, leaving 101 sending times that contribute at least one radio value. Each carrier contributes at most one historical value per metric and sending time; separate SMS and WhatsApp receipts at an April location do not duplicate that value. A missing component is omitted only from its carrier--metric distribution, so these distributions have different denominators. Archived timestamp transcriptions are used to link sending records, not to reconstruct a latency distribution; the field team's confirmation of receipt within 5~s is reported separately in Section~\ref{sec:case:drivetest}.

For each carrier $c$ and radio metric $M\in\{P,Q,S\}$, let $\mathcal B_{cM}$ denote receipt-linked sending times with an available historical spatially matched value of $M$ for that carrier. Here $M_i^c$ is the matched 2025 radio value, not the radio condition at the moment of receipt. The baseline reference is the minimum of these historical values:
\begin{equation}\label{eq:floor}
F_M^c=\min_{i\in\mathcal B_{cM}} M_i^c.
\end{equation}
The sensitivity analysis uses the 10th percentile of the same historical distribution as a stricter reference:
\begin{equation}
G_M^c=Q_{0.10}\{M_i^c:i\in\mathcal B_{cM}\}.
\end{equation}
Table~\ref{tab:carrier_floor} reports both references. Carrier--metric denominators reflect the availability of historical lookup values and are reported in Section~\ref{sec:case:drivetest}.

\begin{table}[pos=!htbp]
\centering
\caption{Historical carrier-specific radio references spatially associated with receipt locations. Each cell gives the minimum / $P_{10}$ of the available historical values.}
\label{tab:carrier_floor}
\small
\begin{tabular}{@{}lccc@{}}
\toprule
Carrier & RSRP (dBm) & RSRQ (dB) & SNR (dB) \\
\midrule
T-Mobile & $-121/-113$ & $-18/-16$ & $-20/-1.4$ \\
AT\&T    & $-131/-123.4$ & $-20/-16$ & $-20/-8.1$ \\
Verizon  & $-123/-116.8$ & $-18/-16.4$ & $-9/-4.0$ \\
\bottomrule
\end{tabular}
\end{table}

Each carrier is evaluated with its own references, and an asset must satisfy all three radio metrics on that same carrier. The procedure therefore cannot combine RSRP from one network, RSRQ from another, and SNR from a third. The references are historical spatial associations, not radio conditions observed when messages were received. They reuse the drive-test layer evaluated in the asset screen and therefore provide no independent delivery validation. The receipt sample also contains no failed delivery, precluding estimation of a delivery-success probability or a boundary separating success from failure. The minimum and $P_{10}$ summarize the historical values at receipt-linked locations; neither is an engineering specification or a guaranteed delivery level. Their comparison identifies sensitivity to the reference choice.

\subsection{Asset screening rules}\label{sec:method:metrics}

The asset connectivity screen and the proposed DMS review tree produce separate outputs. The connectivity screen records whether local radio evidence is unavailable, below the baseline reference, conditional, or robust. The review tree is a deterministic rule, not a fitted model. It combines the baseline connectivity result with inventory fields to assign the next review category.

For each asset $a$, radio conditions are summarized separately for each carrier. Let $\Omega(a)$ denote the retained complete drive-test observations inside the asset footprint. For carrier $c$ and metric $M$, the screen uses the 10th percentile of those observations:
\begin{equation}\label{eq:harmetric}
L_{a,M}^c=Q_{0.10}\{M_j^c:j\in\Omega(a)\}.
\end{equation}
This lower-tail value represents the weaker part of the local measurements, so a few strong observations cannot hide weaker conditions nearby. An asset passes the baseline connectivity screen when at least one carrier meets all three of its own references:
\begin{equation}\label{eq:harpass}
L_{a,P}^c\ge F_P^c,\qquad
L_{a,Q}^c\ge F_Q^c,\qquad
L_{a,S}^c\ge F_S^c.
\end{equation}
The same carrier must satisfy all three inequalities. The sensitivity analysis repeats the test with $G_M^c$ in place of $F_M^c$. Passing this radio screen does not approve replacement of a HAR or DMS. It only advances the asset to a review of whether mobile messaging provides the same traveler-information function and whether the proposed change preserves necessary backup capacity.

The DMS review tree uses five inputs: the baseline screen result, beacon status, ownership, rural designation, and display size. A DMS passes the baseline test when at least one carrier simultaneously meets its RSRP, RSRQ, and SNR references within the 0.5-mi footprint, as defined by equation~\eqref{eq:harpass}. Failure by another carrier does not reverse that pass. A measured sign fails when no carrier meets all three references, while a sign without a retained complete local radio triplet remains unevaluated.

Beacon status and ownership come directly from the inventory. Rural status combines the USDA non-metro classification with the Rural Maryland Council designation. Display size is derived from the character matrix: three-row matrices with 14, 15, 18, 20, or 21 characters are Large; $2\times15$, $3\times8$, and $3\times11$ matrices are Medium; and $0\times0$, blank, or null records are Static. The fields serve as visible implementation proxies. Ownership identifies a separate operating and approval context, an active beacon records an additional roadside warning feature, rurality distinguishes the operating context, and display size describes the installed sign configuration. The fields do not directly measure message history, audience reach, or safety function; agency staff examine those records during service-equivalence review.

The tree applies its branches in a fixed order. A Static record without an active beacon first receives \emph{Inventory review}, even when retained local drive-test measurements are unavailable. Any remaining sign without a retained complete local radio triplet remains unevaluated. Among measured signs, failure of the baseline screen produces Retain for a rural sign and Modernize for an urban sign. For a measured sign that passes, MDTA ownership produces Retain before the tree considers the other attributes. A rural sign with an active beacon also receives Retain. Among the remaining signs, an urban Medium or Static display receives Decommission, and all other signs receive Convert for review of a lower-cost roadside treatment. The formal Decommission label refers the sign to decommissioning review; it does not record an approved removal.

The connectivity evidence state and the DMS review category remain separate in the results. The sensitivity test determines whether a baseline pass is robust or conditional, but it does not replace the category assigned by the baseline tree.

\subsection{Operational constraints and sensitivity}\label{sec:method:sensitivity}

For each asset, reviewers evaluate the messages displayed or broadcast at that location. Recurring travel-time information, bridge closures, weather warnings, and evacuation orders serve different audiences and require different amounts of advance notice. The field team's confirmation that test messages arrived within 5~s describes those tests; reviewers must still compare production-system timing with travel time to the final feasible diversion point. They also examine message content and audience reach, confirm agency authority, and document an independent delivery path for safety-critical messages \citep{dudek2004changeable, bean2015study, olson2024decade, leveson2012engineering}. A missing requirement produces a hold or redesign result rather than being offset by evidence in another category.

Two sensitivity analyses examine whether the asset results depend on the historical reference or the HAR footprint. First, the screen is repeated using each carrier's historical-sample \(P_{10}\) reference instead of its minimum. An asset that passes the baseline test but fails this stricter test receives a conditional evidence state. Second, the HAR analysis is repeated using the agency-provided 5-mi outer ring instead of the 3-mi baseline ring. A HAR that has no retained complete observations within 3~mi but becomes evaluable within 5~mi is reported only in the sensitivity results and is not added to the baseline population.

\subsection{Evidence states and allowable interpretation}\label{sec:method:evidence}

Each asset receives both an evidence state and a review category. The evidence state identifies whether retained complete local measurements are absent, below the baseline reference, conditional, or robust. Robust denotes stability under the executed screening variants, not verified mobile-delivery reliability. The review category identifies the branch assigned by the DMS tree. Table~\ref{tab:evidence_states} specifies the evidence required for each state and the corresponding next step. Even an asset with a robust screen result must complete all three agency reviews before an implementation decision is approved.

\begin{table}[pos=!htbp]
\centering
\caption{Evidence states, empirical definitions, and required next steps.}
\label{tab:evidence_states}
\small
\begin{tabular}{@{}p{0.15\linewidth}p{0.34\linewidth}p{0.42\linewidth}@{}}
\toprule
Evidence state & Empirical basis & Required next step \\
\midrule
Unmeasured & No retained complete radio triplet in the specified footprint & Keep the asset unevaluated; recover missing fields or collect local measurements as appropriate. \\
Below baseline reference & Footprint evidence exists but no carrier's lower-tail values meet all three of its baseline references & Preserve or improve a roadside path and investigate the weak component, campaign conditions, and feasible mobile improvements. \\
Conditional pass & A same-carrier baseline pass is obtained but the asset falls below the carrier-specific $P_{10}$ reference or changes between the 3- and 5-mi HAR rings & Record sensitivity-dependent status and collect repeated or denser measurements before agency review. \\
Robust screen pass & The asset clears the executed same-carrier connectivity variants and an inventory branch produces a label & Advance the asset to message, audience, timing, authority, redundancy, condition, and implementation review. \\
Validated service decision & The quantitative screen and all three agency reviews are complete, with an approved implementation record & Record the approved Retain, Modernize, Convert, or removal decision. \\
\bottomrule
\end{tabular}
\end{table}

Each DMS receives two separate results. Drive-test measurements determine whether its connectivity evidence is robust, conditional, below the baseline reference, or insufficient. The review tree separately uses the baseline connectivity result and inventory fields to assign Inventory review, Retain, Modernize, Convert, or Decommission.

Agency staff evaluate evidence adequacy, service equivalence, and implementation resilience separately. A strong connectivity result does not compensate for missing information about audience reach, message timing, or backup communication. If any review cannot be completed, the agency does not approve decommissioning and records the additional evidence or temporary treatment required before reconsidering the asset.

\subsection{Agency review}\label{sec:method:review}

After the connectivity screen, agency staff review three aspects of each asset: whether the available measurements are sufficient, whether the proposed replacement provides the same traveler-information function, and whether the replacement can be implemented with adequate backup and monitoring. Table~\ref{tab:review_gates} summarizes the evidence examined in each review. An asset-specific decision is approved only when the available evidence supports all three aspects.

\begin{table}[pos=!htbp]
\centering
\caption{Proposed agency reviews required after the asset screen; these reviews were not completed in this study.}
\label{tab:review_gates}
\small
\begin{tabular}{@{}p{0.18\linewidth}p{0.45\linewidth}p{0.29\linewidth}@{}}
\toprule
Review & Information examined & Review outcome \\
\midrule
Evidence adequacy & Location, date, density, missing fields, and carrier coverage of drive-test data; the historical nature and spatial matching of the receipt-associated radio values & Determine whether the available evidence is sufficient or additional local measurements are needed \\
Service equivalence & Messages and hazards served by the asset; delivery timing, audience reach, accessibility, authority, and notification before the last feasible decision point & Determine whether the proposed replacement provides the required traveler-information function \\
Implementation resilience & Backup communication, asset condition, ownership, conversion design, reversibility, monitoring, and agency approval & Determine whether the replacement can be implemented and monitored with an adequate contingency \\
\bottomrule
\end{tabular}
\end{table}

The final record reports the connectivity evidence, sensitivity result, preliminary review category, agency decision, and any reason for changing that category. New drive tests or changes in the roadside asset can trigger another review without removing the evidence used for the previous decision.

\subsection{Policy-scenario comparison and decision-rule audit}\label{sec:method:audit}

The decision audit compares analyst-defined screening scenarios on the same 234-asset inventory. These scenarios are controlled alternatives, not descriptions of adopted agency policies. The inventory-only scenario bypasses the measurement and connectivity checks while retaining the inventory branches. The connectivity-only scenario identifies assets with a same-carrier baseline pass without assigning an implementation category. The baseline scenario applies the proposed tree. An evidence-gated scenario additionally places conditional Retain, Convert, and Decommission records on hold for further measurement. Inventory-review records retain their existing status. The comparison reports category counts and asset-level transitions; it does not estimate safety gains, cost savings, or classification accuracy against an external standard.

A condition-only DMS comparison cannot be calculated from the available inventory because it does not contain DMS physical-condition ratings or maintenance costs. Ownership, beacon status, and display size are not substituted for those variables. The HAR inventory does contain recorded communications status, so the analysis reports which HAR assets a condition review would flag, separately from their mobile-connectivity results.

One-at-a-time rule variants test the dependence of the DMS labels on implementation assumptions. The ownership-removal variant omits the MDTA retention branch. The branch-order variant places the urban Medium/Static rule before the ownership branch. The beacon variant gives all measured, baseline-passing signs with active beacons the retention treatment otherwise reserved for rural signs. The unknown-field variant sends blank display sizes and blank or null beacon values to Inventory review before applying other branches; an explicit $0\times0$ record retains its baseline Static coding. These alternatives test the consequences of the coding and rule choices without asserting that any alternative is agency-approved.

Three measurement variants replace the minimum historical reference with the historical-sample $P_{10}$, require baseline support on all three carriers, or change the DMS radius from 0.5~mi to 0.25 or 1~mi. The all-carrier variant leaves a sign unevaluated if any carrier lacks retained complete observations, except for records already routed to Inventory review. When all carriers have complete observations but the all-carrier criterion fails, the existing rural Retain/urban Modernize branch applies. The stricter-reference variant changes the tree's screening input; this differs from the evidence-gated scenario, which preserves the baseline category and adds a hold. All other inputs remain fixed. The exported transition table identifies every changed asset.

The three baseline decommissioning-review records are examined individually using inventory identifiers, location, carrier sample counts, spatial-cell counts, and the stricter radio reference. The audit also calculates the fraction of each carrier's local observations that meets all three component references in the same observation:
\begin{equation}\label{eq:joint_fraction}
J_{a,c}(H)=\frac{1}{n_{a,c}}\sum_{j\in\Omega_c(a)}\mathbf{1}\{P_j^c\ge H_P^c,\ Q_j^c\ge H_Q^c,\ S_j^c\ge H_S^c\},
\end{equation}
where $\Omega_c(a)$ contains the retained complete footprint observations for carrier $c$, $n_{a,c}=|\Omega_c(a)|$, and $H$ is either the minimum or stricter historical reference. The fraction is undefined when $n_{a,c}=0$. It describes joint exceedance of historical radio references, not message-delivery probability. Separate marginal $P_{10}$ tests do not imply that 90\% of observations jointly satisfy all three references. Spatial cells count sampled locations using the existing route-specific grid; they are not independent statistical replicates. Shared observation identifiers are counted for overlapping asset footprints.

For each case, the decision record distinguishes historical radio evidence, inventory assumptions, and unresolved service requirements. Completion requires the actual message history and beacon function, the served audience, timing before the last useful decision point, independent backup, condition and lifecycle-cost records, and an authorized agency judgment. A missing record remains unresolved. The case audit therefore evaluates the basis for referral and identifies what would be needed for a final decision; it does not certify service replacement.

OpenAI Codex (OpenAI) assisted the development and revision of deterministic analysis code, consistency checks, and interpretation of the outputs. Calculations used the archived measurements without AI-generated or imputed radio values, and archived receipt timestamps were not rewritten. The authors retain responsibility for reviewing the assumptions, code, and results and for all reported interpretations.

\section{Maryland case study}
\label{sec:casestudy}

\subsection{Study area and assets}\label{sec:case:area}

\begin{figure}[pos=!htbp]
\centering
\includegraphics[width=0.94\linewidth]{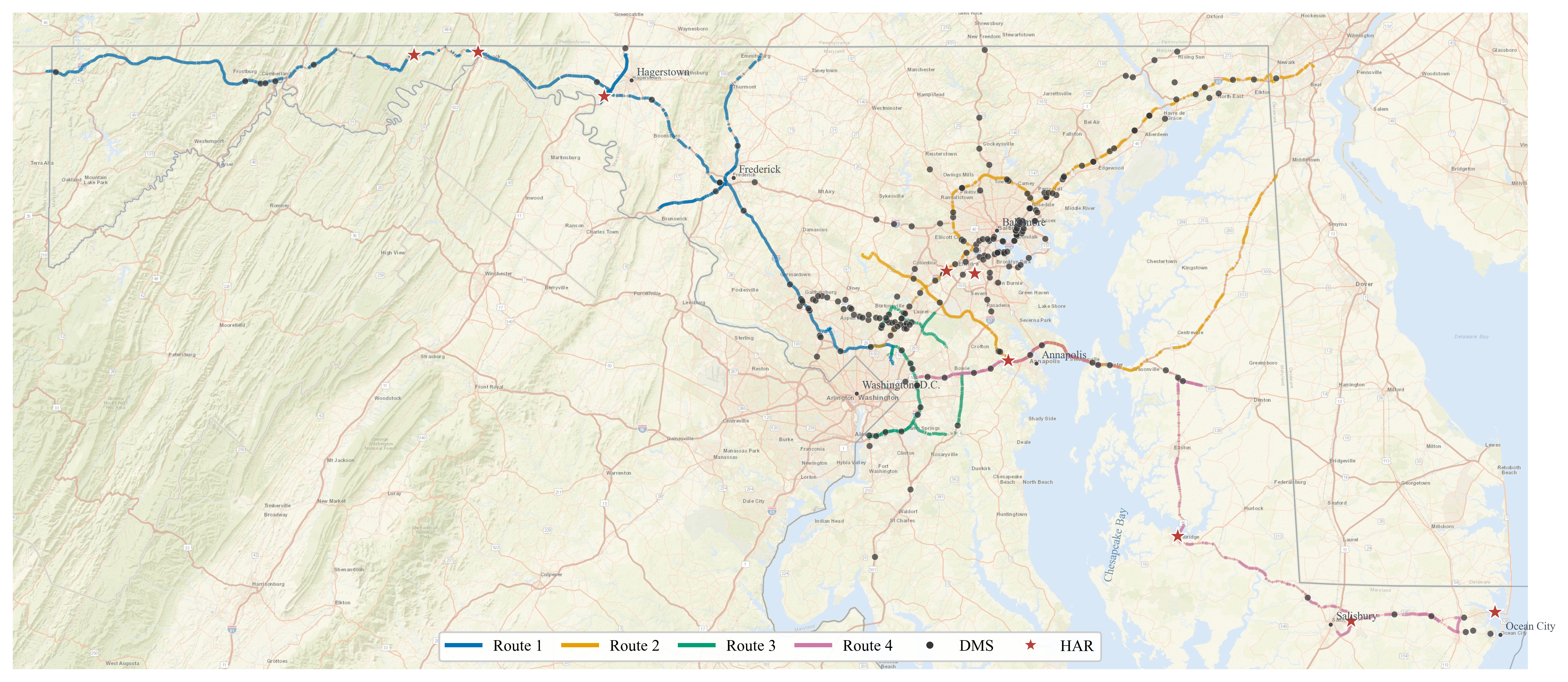}
\caption{The four Maryland drive-test corridors (C1--C4, shown as Routes 1--4) and the locations of the HAR and DMS assets. The muted outline marks the Maryland state boundary; the ArcGIS basemap and selected city and waterbody labels provide geographic context.}
\label{fig:layout}
\end{figure}

MDSHA and the Coordinated Highways Action Response Team (CHART) distribute traveler information through roadside assets, 511 services, mobile alerts, and third-party feeds \citep{chang2009performance}. The case study covers four operating environments: C1, the I-70/I-68 corridor through mountainous Western Maryland; C2, a loop across the Bay Bridge and upper Eastern Shore with a return on I-95; C3, the exurban fringe of southern Maryland; and C4, the lower Eastern Shore and Ocean City corridor. These corridors represent mountain terrain, constrained diversion points, suburban--rural transitions, and seasonal visitor travel, respectively. Figure~\ref{fig:layout} shows these corridors as Routes 1--4 and locates the nine HAR transmitters and 234 DMS included in the inventory.

C1 follows I-270, I-70, and I-68 from the Washington region through Frederick and Hancock to Garrett County. Ridge-and-valley terrain, long distances between interchanges, winter weather, and recreational travel make widespread cellular gaps plausible on this corridor. Drive testing was conducted on October~22, 2025.

C2 crosses the Chesapeake Bay Bridge on US-50/US-301, continues north through the upper Eastern Shore to Cecil County, and returns on I-95. The route includes rural two-lane roads, toll facilities, and urban freeways. Drivers approaching the Bay Bridge must receive a warning before the last exit that provides a practical alternative route. A mobile notification received after that exit cannot provide the same advance warning as an upstream HAR or DMS. Drive testing was conducted on November~27--28, 2025.

C3 follows MD-4 and US-301 through the exurban portion of southern Prince George's County. This route was included to examine mobile coverage on partly rural roads near the Washington metropolitan area. Drive testing was conducted on December~1, 2025. 

C4 follows US-50 through Easton, Cambridge, and Salisbury toward Ocean City. Its seasonal visitors, bridge approaches, and limited detour options make audience reach and application enrollment particularly relevant. Drive testing was conducted on October~14, 2025.

The HAR inventory contains nine transmitters. Four (HAR-2, HAR-3, HAR-4, and HAR-6) were marked as communication failures in the inventory snapshot; this status is independent of the cellular screen. Seven units have retained complete drive-test observations within the baseline 3-mi footprint. HAR-7 and HAR-8 do not and remain unevaluated in the baseline analysis.

\begin{table}[pos=!htbp]
\centering
\caption{Maryland HAR inventory. Status is the recorded roadside communications status and is tracked separately from the mobile connectivity result.}
\label{tab:har_inventory}
\small
\begin{tabular}{@{}llll@{}}
\toprule
ID & Location & Corridor & Status \\
\midrule
HAR-1 & I-95 at MD 100 & C2 & OK \\
HAR-2 & I-81 at US 11 & C1 & Comm. failure \\
HAR-3 & US 522 at I-70 & C1 & Comm. failure \\
HAR-4 & I-68 at Orleans Rd & C1 & Comm. failure \\
HAR-5 & US 50 at MD 16 & C4 & OK \\
HAR-6 & US 50 at US 13 & C4 & Comm. failure \\
HAR-7 & MD 90, St. Martins Neck & C4 & OK \\
HAR-8 & MD 170 at BWI & Off-corridor & OK \\
HAR-9 & I-97 at US 50 & C2 & OK \\
\bottomrule
\end{tabular}
\end{table}

The inventory also records AM frequency, station call sign, communications mode, and SHA-provided 3- and 5-mi broadcast rings. Several units use POTS or legacy cellular backhaul. The 3-mi inner ring defines the baseline measurement footprint, and the 5-mi outer ring is used for sensitivity analysis. Recorded communication failure and the mobile connectivity result are retained as separate fields for subsequent agency review.

The DMS inventory contains 234 signs: 105 MDSHA/CHART assets and 129 Maryland Transportation Authority (MDTA) assets. Coordinates are available for every sign, display size for 92\%, and active-beacon status for 208 signs. The retained-data spatial assignment associates 120 signs with the four driven corridors; 105 of these have at least one complete radio triplet within their evaluation footprints (Table~\ref{tab:dms_corridor}). The other 114 signs receive the off-corridor designation, including 72 MDTA signs near Baltimore tunnels and beltways. These signs remain unevaluated unless the inventory-only rule applies.

\begin{table}[pos=!htbp]
\centering
\caption{DMS inventory and availability of drive-test evidence.}
\label{tab:dms_corridor}
\small
\begin{tabular}{@{}lrrrr@{}}
\toprule
Corridor & DMS & MDSHA & MDTA & With drive-test \\
\midrule
C1 Western Maryland       & 23  & 17  & 6  & 21 \\
C2 Bay--upper Shore--NE   & 54  & 22  & 32 & 47 \\
C3 Southern Maryland      & 27  & 12  & 15 & 22 \\
C4 Lower Eastern Shore    & 16  & 12  & 4  & 15 \\
Off-corridor              & 114 & 42  & 72 & 0 \\
\midrule
Total                     & 234 & 105 & 129 & 105 \\
\bottomrule
\end{tabular}
\end{table}

\subsection{Drive tests and delivery observations}\label{sec:case:drivetest}

Drive testing was conducted along the four corridors between October~14 and December~1, 2025, using three commercial phones, one for each carrier. The resulting dataset contains 160{,}232 GPS-referenced observations: 44{,}268 on C1, 43{,}190 on C2, 19{,}869 on C3, and 52{,}905 on C4. After observations without a complete RSRP, RSRQ, and SNR triplet were removed, 128{,}476 observations remained for the main analysis: 47{,}242 for T-Mobile, 34{,}079 for AT\&T, and 47{,}155 for Verizon. Complete triplets yielding a zero score under Eq.~\eqref{eq:oe} were retained; 31{,}756 incomplete observations were excluded. This complete-triplet rule does not establish that all no-service events were preserved. The separate score-only sensitivity analysis described in Section~\ref{sec:method:scoring} reinstates 3{,}287 incomplete records whose available fields already determine $O=0$, without imputing missing radio values.

Each retained observation includes the route, carrier, timestamp, and GPS coordinates. Direct carrier comparisons use spatial matching rather than assuming equivalent observations from identical timestamps. C4 contains different numbers of observations for the three carriers, making spatial matching necessary for its comparisons. Horizontal-accuracy and vehicle-motion fields were retained for data-quality checks. All three phones were mounted in the same vehicle to reduce differences in device position. The data represent conditions during the study period and should not be interpreted as long-term measures of network performance.

\begin{figure}[pos=!htbp]
\centering
\includegraphics[width=0.98\linewidth]{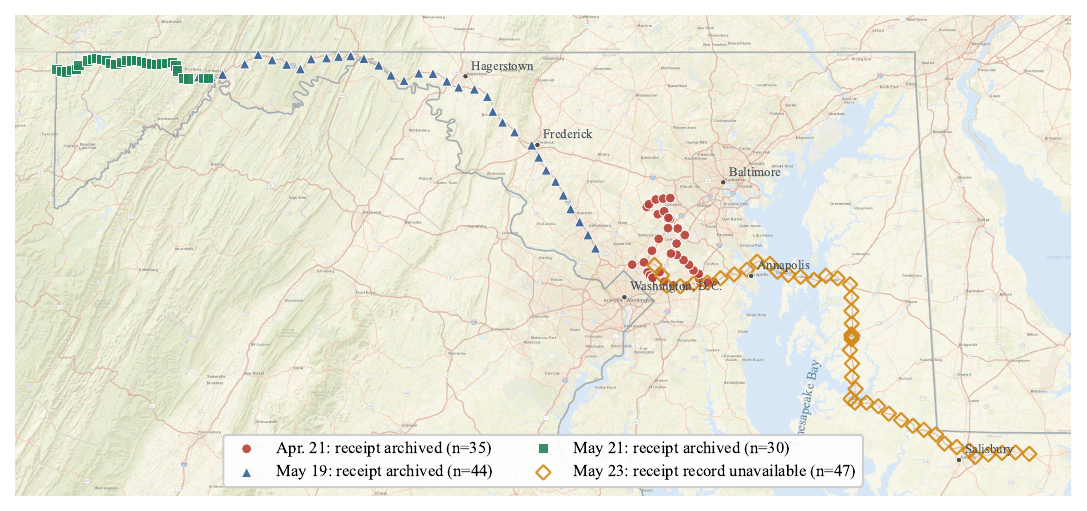}
\caption{All 156 georeferenced message-sending locations. Filled symbols identify 109 locations with archived receipt evidence; 101 have at least one available radio component from the historical spatial lookup. Open diamonds identify the 47 May~23 Eastern Shore locations for which receipt records are unavailable in the local archive. The muted outline marks the Maryland state boundary. Boundary source: U.S. Census Bureau TIGERweb. Basemap sources: Esri, HERE, Garmin, USGS, FAO, NPS, NRCAN, GeoBase, IGN, Kadaster NL, Ordnance Survey, OpenStreetMap contributors, and the GIS User Community.}
\label{fig:field_test_locations}
\end{figure}

Follow-up field tests in 2026 documented message receipt and georeferenced sending times; the radio fields subsequently associated with these locations came from the 2025 drive-test layer. On April~21, 2026, fixed messages were sent from 35 locations through SMS and WhatsApp, yielding 210 documented successful carrier--channel deliveries. Three additional sessions were conducted on May~19, 21, and 23. Their GPS schedule contains 121 sending locations: 44 on May~19, 30 on May~21, and 47 on May~23. The May receipt archive contains 100 sending times from May~19 and 21, but no May~23 receipt records. After the timestamps were normalized to the minute, 74 archived sending times matched the GPS schedule and contained receipt fields for all three carriers. Figure~\ref{fig:field_test_locations} maps all 156 georeferenced field-test locations and distinguishes the 47 Eastern Shore locations without receipt records in the local archive. The 35 April and 74 timestamp-matched May locations provide 109 receipt-linked sending times and 432 documented successful deliveries. Eight April times have no historical radio value for any carrier; thus 101 sending times contribute to at least one historical carrier--metric reference. No failed delivery is present in the retained receipt sample.

\begin{table}[pos=!htbp]
\centering
\caption{Georeferenced field-test locations and the subset linked to archived receipts. Of the 109 receipt-linked times, 101 contribute at least one historical radio value.}
\label{tab:push_campaigns}
\small
\begin{tabular}{@{}lrrr@{}}
\toprule
Date & Georeferenced times & Receipt-linked times & Successful deliveries \\
\midrule
April 21 & 35 & 35 & 210 \\
May 19 & 44 & 44 & 132 \\
May 21 & 30 & 30 & 90 \\
May 23 & 47 & 0 & --- \\
\midrule
Total mapped & 156 & 109 & 432 \\
\bottomrule
\end{tabular}
\end{table}

The field team confirmed that every tested message was received within 5~s of transmission. This statement reports the team's field observation; the archived timestamp transcriptions are not used to estimate a latency distribution. Sending-time linkage identifies the retained receipt records, while spatial lookup supplies the associated historical radio values. The 47 May~23 locations document Eastern Shore field coverage but do not enter the historical reference sample because their receipt records are unavailable in the local archive.

The reference denominators are the available historical values for each carrier and metric, not 109 complete triplets. The retained locations provide 98 RSRP values, 98 RSRQ values, and 97 SNR values for T-Mobile; 100, 75, and 100 values, respectively, for AT\&T; and 97 values for each Verizon metric. The eight April locations with no radio values contribute to the receipt count but to none of these distributions. Additional component-level missingness accounts for the remaining differences.

The carrier references are pooled across corridors because the receipt-linked locations are unevenly distributed and too sparse for route-specific reference distributions. The minimum retains the weakest available historical value in each carrier--metric sample, while the $P_{10}$ reduces the influence of extreme values. These distributions describe historical spatial associations with documented receipt locations. They do not measure radio conditions at receipt, and their reuse of the 2025 drive-test layer provides no independent validation of the asset screen.

\FloatBarrier
\section{Results}\label{sec:results}

\subsection{Carrier and corridor connectivity}\label{sec:results:baseline}

The carrier results vary markedly across the four corridors (Table~\ref{tab:carrier_summary}). Across all scored observations, T-Mobile has the highest mean and median connectivity score. The aggregate ranking does not hold on every route: Verizon has the highest mean on C2, while T-Mobile leads on C1, C3, and C4. The low-score share also changes by route and carrier, reaching 57.0\% for Verizon on C1 and 66.2\% for AT\&T on C2. These reversals show why a substitution decision cannot be based on a single statewide carrier average.

\begin{table}[pos=!htbp]
\centering
\caption{Drive-test connectivity scores by carrier, using records with complete RSRP, RSRQ, and SNR values. Zero scores in this complete-record sample are retained.}
\label{tab:carrier_summary}
\small
\begin{tabular}{@{}lrrrrrrr@{}}
\toprule
Carrier & $n$ & Mean & Median & $P_{25}$ & $P_{75}$ & \% $<10$ & \% $\ge70$ \\
\midrule
T-Mobile & 47{,}242 & 46.5 & 50.5 & 20.9 & 70.7 & 21.8 & 26.0 \\
AT\&T    & 34{,}079 & 34.8 & 36.1 & 0.0  & 59.9 & 35.0 & 16.4 \\
Verizon  & 47{,}155 & 39.9 & 40.5 & 0.0  & 68.2 & 32.2 & 23.7 \\
\bottomrule
\end{tabular}
\end{table}

\begin{figure}[pos=!htbp]
\centering
\includegraphics[width=\linewidth]{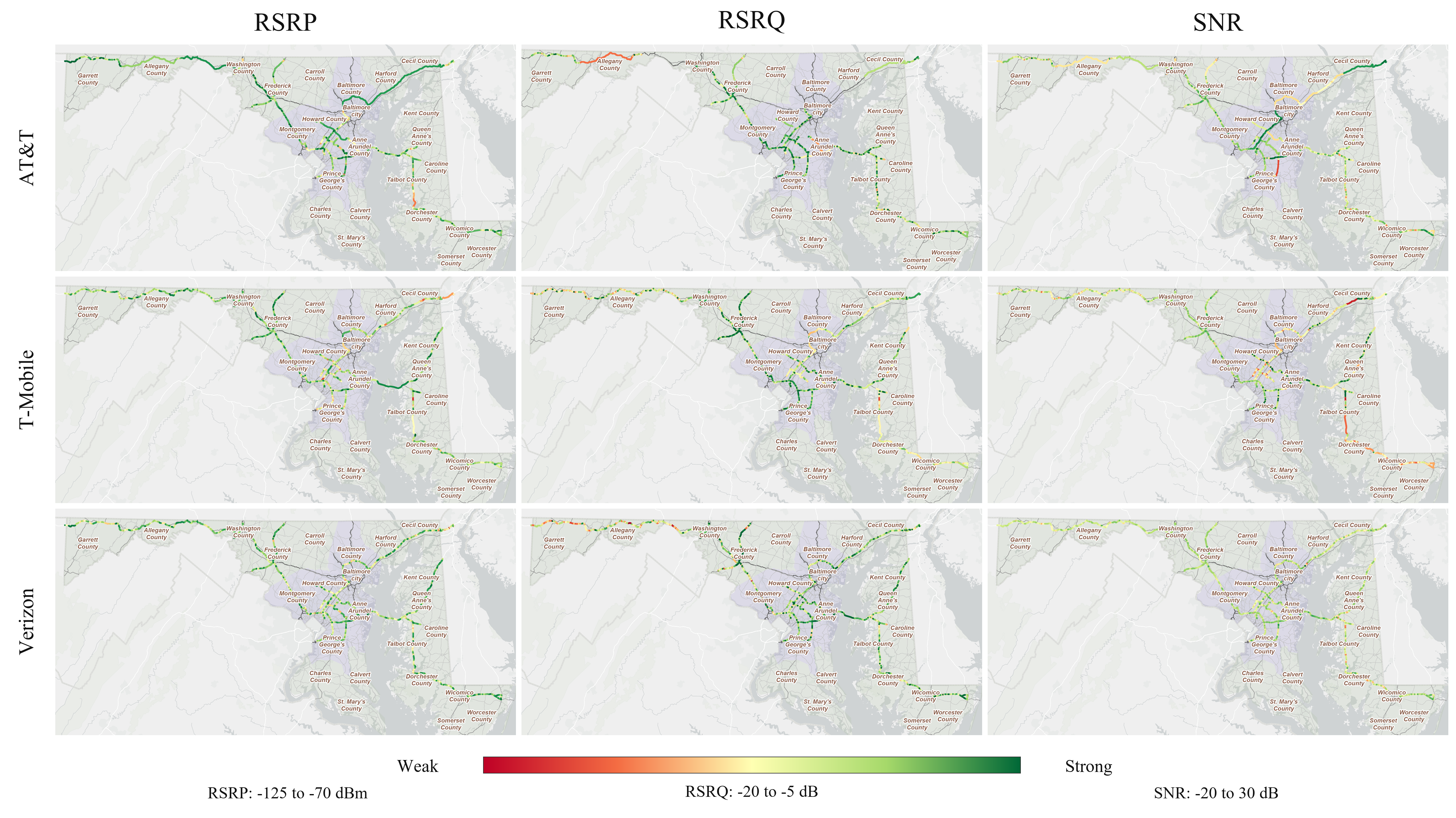}
\caption{RSRP, RSRQ, and SNR measured on the four corridors. Each metric uses the same weak-to-strong color ramp for all three carriers; the metric-specific ranges are shown below the maps.}
\label{fig:signal_strength_grid}
\end{figure}

\begin{figure}[pos=!htbp]
\centering
\includegraphics[width=0.94\linewidth]{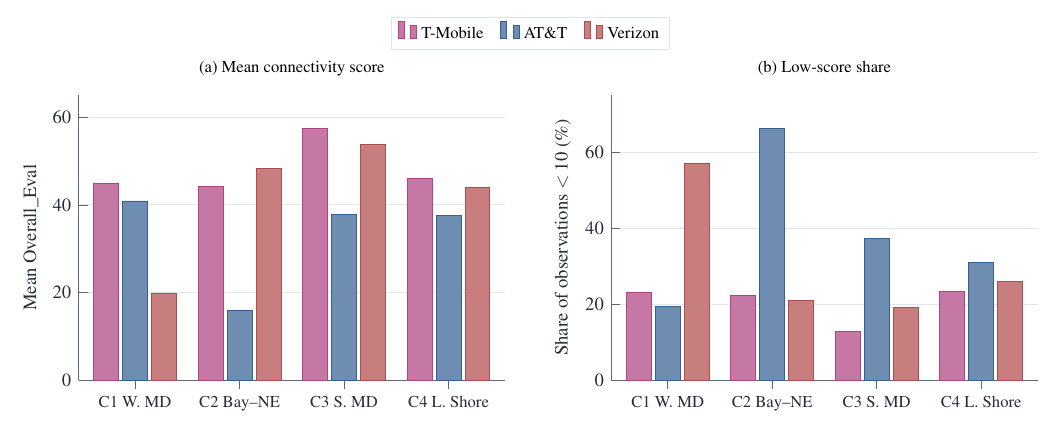}
\caption{Mean connectivity score and share of observations with a score below 10, by corridor and carrier.}
\label{fig:route_carrier}
\end{figure}

\begin{table}[pos=!htbp]
\centering
\caption{Route--carrier connectivity-score statistics.}
\label{tab:route_carrier_summary}
\small
\begin{tabular}{@{}llrrrrr@{}}
\toprule
Corridor & Carrier & $n$ & Mean & Median & \% $<10$ & \% $\ge70$ \\
\midrule
C1 Western Maryland & T-Mobile & 14{,}735 & 44.9 & 47.6 & 23.1 & 25.5 \\
 & AT\&T & 10{,}147 & 40.9 & 41.5 & 19.4 & 18.3 \\
 & Verizon & 12{,}967 & 19.8 & 0.0 & 57.0 & 4.3 \\
\midrule
C2 Bay--upper Shore--NE & T-Mobile & 11{,}502 & 44.2 & 45.2 & 22.3 & 27.1 \\
 & AT\&T & 6{,}117 & 16.0 & 0.0 & 66.2 & 3.3 \\
 & Verizon & 12{,}615 & 48.3 & 52.3 & 21.0 & 31.2 \\
\midrule
C3 Southern Maryland & T-Mobile & 5{,}328 & 57.4 & 61.3 & 12.7 & 37.9 \\
 & AT\&T & 6{,}206 & 37.9 & 41.4 & 37.2 & 21.5 \\
 & Verizon & 6{,}622 & 53.9 & 59.8 & 19.1 & 39.6 \\
\midrule
C4 Lower Eastern Shore & T-Mobile & 15{,}677 & 46.1 & 52.6 & 23.3 & 21.6 \\
 & AT\&T & 11{,}609 & 37.6 & 37.8 & 31.0 & 19.0 \\
 & Verizon & 14{,}951 & 44.0 & 44.1 & 26.0 & 27.3 \\
\bottomrule
\end{tabular}
\end{table}

The direct cross-carrier analysis uses 3{,}723 route-specific grid cells constructed from T-Mobile observations with AT\&T and Verizon matches within 5~m. At the selected score threshold of 30, T-Mobile meets the criterion in 73.1\% of cells, Verizon in 57.1\%, and AT\&T in 54.0\%. At least one carrier meets the criterion in 92.8\% of cells; all three fall below it in 7.2\%. The all-carrier low-score share is highest on C1 (12.1\%), followed by C2 (5.4\%), C4 (4.7\%), and C3 (2.0\%). The best-of-three value is an upper bound on available infrastructure, not the coverage available to a traveler subscribed to one carrier.

\begin{figure}[pos=!htbp]
\centering
\includegraphics[width=0.97\linewidth]{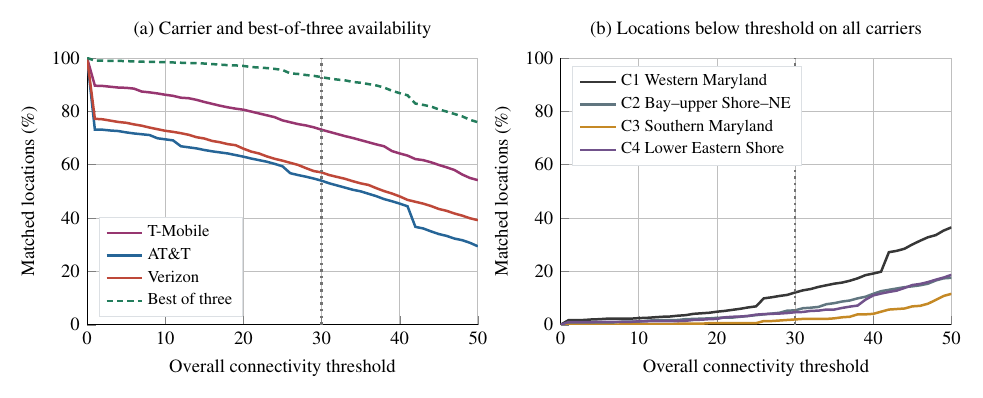}
\caption{Sensitivity of matched-location results to the connectivity-score threshold. The dotted line marks the value of 30 used for the main corridor comparison.}
\label{fig:oe_threshold}
\end{figure}

Figure~\ref{fig:oe_threshold} shows that the conclusion depends on what the score threshold is intended to represent. Raising it from 30 to 50 reduces the best-of-three share from 92.8\% to 75.9\% and increases the all-carrier low-score share from 7.2\% to 24.1\%. The threshold is therefore used as a transparent screening setting rather than as a guarantee of message delivery. C1 remains the corridor with the largest all-carrier low-score share throughout the upper part of the tested range.

The complete-record filter also affects the corridor results. Among the 31{,}756 excluded records, 3{,}287 have an available RSRP or SNR value that already determines a zero score under Eq.~\eqref{eq:oe}: 3{,}115 AT\&T records and 172 T-Mobile records. Including these records for a separate corridor sensitivity analysis, without imputing any missing radio component, reduces C1 AT\&T's mean score from 40.9 to 32.4 and increases its share below 10 from 19.4\% to 36.1\%. Repeating the same spatial matching and grid aggregation yields 4{,}113 complete carrier-score cells; the best-of-three share at 30 becomes 91.5\%, and the all-carrier low-score share becomes 8.5\%. This change reflects both additional zero scores and a changed matched-cell population. Tables~\ref{tab:carrier_summary} and~\ref{tab:route_carrier_summary} and Figures~\ref{fig:route_carrier} and~\ref{fig:oe_threshold} retain the complete-record main-analysis definition. The sensitivity result shows that excluding incomplete records can understate low-score exposure; it neither identifies every no-service event nor supplies the missing raw components required for the asset screen.

\FloatBarrier
\subsection{Spatially associated radio references and HAR screening}\label{sec:results:har}

The GPS schedule provides 156 georeferenced message-sending locations, including 47 from the May~23 Eastern Shore session (Figure~\ref{fig:field_test_locations}). Locally archived receipt records cover 109 of these locations: 35 from the April test and 74 from the May~19 and 21 sessions. All 432 carrier--channel deliveries in this retained receipt sample were documented as successfully received; the sample contains no failed delivery. Of the 109 locations, 101 have at least one spatially assigned historical radio value, while eight April locations have none. Because no May~23 receipt records are available in the local archive, those Eastern Shore locations document field coverage but do not contribute to the exploratory references in Table~\ref{tab:carrier_floor}. The field team confirms that the tested messages were received within 5~s of sending; this statement is based on the confirmed field observation, not on delay calculations from the transcribed timestamp fields.

The minimum and $P_{10}$ differ most for RSRP and SNR. For example, T-Mobile's RSRP reference changes from $-121$ to $-113$~dBm, while its SNR reference changes from $-20$ to $-1.4$~dB. The minimum preserves the weakest historical value assigned to the receipt-linked test locations; it is not the weakest signal measured during successful receipt. The $P_{10}$ reference reduces the effect of extreme historical values. Both are exploratory comparison settings, not engineering specifications or promises of future delivery.

\begin{table}[pos=!htbp]
\centering
\caption{HAR screening results. T, A, and V denote T-Mobile, AT\&T, and Verizon.}
\label{tab:har_results}
\small
\begin{tabular}{@{}lcccc@{}}
\toprule
ID & Baseline, 3 mi & $P_{10}$, 3 mi & $P_{10}$, 5 mi & State \\
\midrule
HAR-1 & T/A/V & T/V & T/V & Robust \\
HAR-2 & T/A/V & --- & --- & Conditional \\
HAR-3 & T/V & T/V & T & Robust \\
HAR-4 & T/A/V & T/A & A & Robust \\
HAR-5 & A/V & A/V & A/V & Robust \\
HAR-6 & T/A & T/A & T/A & Robust \\
HAR-7 & --- & --- & --- & Insufficient \\
HAR-8 & --- & --- & T/V & Insufficient$^{a}$ \\
HAR-9 & T/A/V & V & --- & Conditional \\
\bottomrule
\multicolumn{5}{l}{\footnotesize $^{a}$No observations occur within 3 mi; support appears only in the 5-mi sensitivity ring.}
\end{tabular}
\end{table}

Seven of the nine HAR sites have observations within 3~mi, and all seven pass the baseline same-carrier test. Five remain supported under both the stricter reference and the 5-mi footprint and are labeled robust. HAR-2 passes only with the baseline minimum. HAR-9 passes the stricter test at 3~mi through Verizon but not when the footprint expands to 5~mi, showing that its result is sensitive to the spatial range. HAR-7 has no observations within either ring. HAR-8 has no 3-mi observations but gains T-Mobile and Verizon support in the 5-mi sensitivity analysis; it remains an evidence gap in the baseline population.

The four HAR sites recorded as roadside communication failures (HAR-2, HAR-3, HAR-4, and HAR-6) all pass the baseline mobile-connectivity screen. This is not a contradiction: the inventory field describes the condition of the roadside HAR system, whereas the screen describes nearby cellular measurements. The two fields answer different questions and both remain in the agency review record.

\FloatBarrier
\subsection{DMS review categories and evidence states}\label{sec:results:dms}

The DMS inventory contains 234 signs. Local observations are available within the 0.5-mi footprint of 105 signs, and every measured footprint passes the baseline same-carrier screen. The baseline connectivity result therefore determines whether a sign can enter the measured branches but does not distinguish the measured signs from one another in this dataset. The inventory branches assign Retain to 58 signs, Convert to 40, and Decommission to three. Four additional measured signs and 14 unmeasured signs enter Inventory review because they are Static records without an active beacon. The remaining 115 unmeasured signs receive Insufficient drive-test coverage. No sign enters Modernize because that branch requires an urban measured sign that fails the baseline screen.

\begin{figure}[pos=!htbp]
\centering
\includegraphics[width=0.95\linewidth]{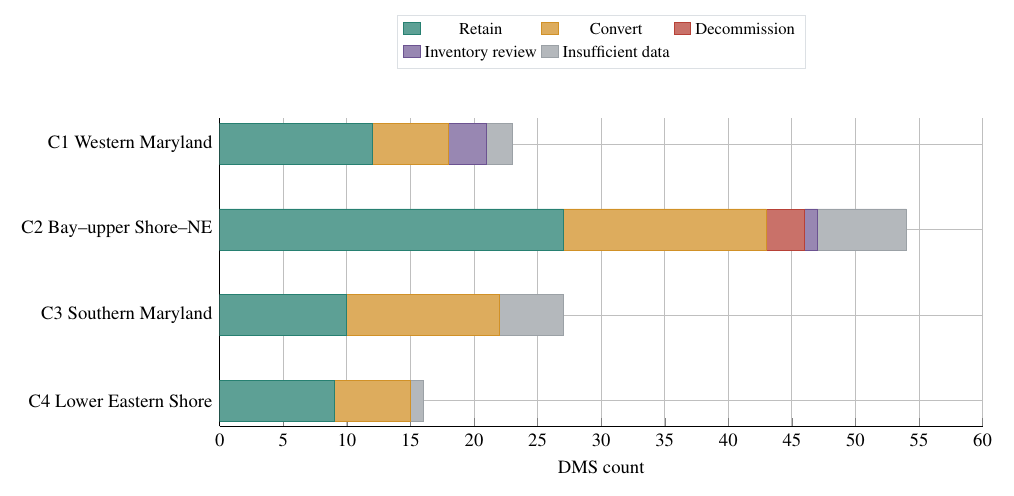}
\caption{DMS review categories on the four driven corridors, including inventory-review and insufficient-evidence records. Off-corridor assets are reported in Table~\ref{tab:dms_results}.}
\label{fig:dms_label}
\end{figure}

\begin{table}[pos=!htbp]
\centering
\caption{DMS review categories produced by the proposed decision tree. The Decommission category denotes referral to decommissioning review.}
\label{tab:dms_results}
\small
\begin{tabular}{@{}lrrrrrrr@{}}
\toprule
Corridor & DMS & Retain & Modernize & Convert & Decommission & Inventory review & Insufficient \\
\midrule
C1 Western Maryland & 23 & 12 & 0 & 6 & 0 & 3 & 2 \\
C2 Bay--upper Shore--NE & 54 & 27 & 0 & 16 & 3 & 1 & 7 \\
C3 Southern Maryland & 27 & 10 & 0 & 12 & 0 & 0 & 5 \\
C4 Lower Eastern Shore & 16 & 9 & 0 & 6 & 0 & 0 & 1 \\
Off-corridor & 114 & 0 & 0 & 0 & 0 & 14 & 100 \\
\midrule
Total & 234 & 58 & 0 & 40 & 3 & 18 & 115 \\
\bottomrule
\end{tabular}
\end{table}

The order of the inventory branches affects the results. The tree assigns an MDTA-owned sign to Retain before evaluating the urban Medium/Static branch. Measured MDTA signs therefore remain in Retain, while the three signs assigned to Decommission are non-MDTA signs on C2. These three records enter decommissioning review rather than an approved removal action.

The stricter spatially associated references divide the 105 measured DMS into 88 robust and 17 conditional connectivity states. Among the 101 signs assigned Retain, Convert, or Decommission, 86 are robust and 15 are conditional; two other conditional signs and two robust signs belong to Inventory review. The 115 signs assigned Insufficient drive-test coverage remain unevaluated rather than connectivity failures. Of these, 100 are off corridor and require a separate campaign before any portfolio-wide inference; the 15 corridor-assigned gaps are the immediate measurement priority.

\FloatBarrier
\subsection{Policy-scenario comparison}\label{sec:results:scenarios}

Table~\ref{tab:policy_scenarios} shows how the evidence requirements change the screening portfolio. Bypassing local measurements assigns an inventory-based action category to 216 assets, compared with 101 under the baseline tree. The additional 115 assignments comprise 84 Retain, 28 Convert, and three Decommission records that lack retained complete local radio measurements. The baseline procedure withholds these assignments; their true service suitability remains unknown. This is a reduction in unsupported assignments under the stated evidence requirement, not a measured reduction in erroneous decisions.

\begin{table}[pos=!htbp]
\centering\small
\caption{Analyst-defined DMS screening scenarios on the same 234 assets. Decommission denotes referral to review. No scenario approves removal.}
\label{tab:policy_scenarios}
\begin{tabular}{@{}lrrrr@{}}
\toprule
Output & Inventory only & Connectivity only & Baseline tree & Evidence gated \\
\midrule
Retain & 142 & --- & 58 & 50 \\
Convert & 68 & --- & 40 & 33 \\
Decommission & 6 & --- & 3 & 3 \\
Inventory review & 18 & --- & 18 & 18 \\
Insufficient local evidence & 0 & 129 & 115 & 115 \\
Hold for further measurement & 0 & --- & 0 & 15 \\
Radio candidate, category unset & --- & 105 & --- & --- \\
\bottomrule
\end{tabular}
\end{table}

Connectivity alone identifies 105 radio candidates, including four records that the baseline tree assigns to Inventory review. It cannot select among retention, conversion, and decommissioning review. Requiring a robust rather than conditional evidence state before advancing a classified asset places eight Retain and seven Convert records on hold, leaving 86 records with robust evidence in the three action-review categories. All three decommissioning-review records retain their radio support. The distinction between the 129 unmeasured assets and the baseline table's 115 Insufficient records arises because 14 unmeasured assets receive inventory-only review.

The HAR condition comparison identifies HAR-2, HAR-3, HAR-4, and HAR-6 for review of recorded communications failures. All four pass the baseline mobile screen, but HAR-2 is conditional under the stricter reference. A condition-based maintenance priority therefore does not identify the same question as mobile substitutability. No comparable condition-only result is reported for DMS because the required condition and cost fields are absent.

\FloatBarrier
\subsection{Sensitivity of the DMS decision rules}\label{sec:results:rules}

Removing the MDTA retention branch changes 38 labels: 20 Retain records become Convert and 18 become Decommission. Moving the urban Medium/Static branch ahead of ownership changes only the latter 18 records. Both variants increase the decommissioning-review count from three to 21. Giving every active beacon the rural retention treatment changes 39 labels, including all three baseline decommissioning-review records. These results identify the assets affected by the institutional assumptions; they do not establish which assumption is preferable.

\begin{table}[pos=!htbp]
\centering\small
\caption{One-at-a-time DMS sensitivity tests. Changed counts compare all 234 labels with the baseline. Inventory review is retained as a distinct category.}
\label{tab:rule_sensitivity}
\begin{tabular}{@{}lrrr@{}}
\toprule
Variant & Changed & Decommission & Insufficient \\
\midrule
Baseline & 0 & 3 & 115 \\
Remove ownership retention & 38 & 21 & 115 \\
Urban Medium/Static before ownership & 18 & 21 & 115 \\
Retain all active-beacon signs after radio pass & 39 & 0 & 115 \\
Hold unknown inventory fields for review & 8 & 3 & 111 \\
Use stricter radio reference in tree & 13 & 3 & 115 \\
Require all three carriers & 36 & 0 & 141 \\
Reduce footprint to 0.25 mi & 5 & 3 & 120 \\
Expand footprint to 1 mi & 15 & 3 & 100 \\
\bottomrule
\end{tabular}
\end{table}

Separating unknown inventory fields changes eight labels: four Convert and four Insufficient records move to Inventory review. The latter transition clarifies the next review task and does not create local radio evidence. Replacing the baseline reference with the stricter reference changes 13 labels to Modernize under the urban failure branch. By contrast, the evidence-gated scenario places conditional records on hold without interpreting a stricter-reference failure as a modernization prescription. Requiring support on all three carriers changes 36 labels and leaves all three baseline decommissioning-review assets unevaluated because their footprints contain no retained AT\&T observations.

Changing the footprint produces a different effect. A 0.25-mi radius removes local evidence for five baseline Retain assets; a 1-mi radius supplies observations for 15 previously Insufficient assets, assigning 11 to Retain and four to Convert. The three baseline decommissioning-review labels remain unchanged under both radii and the stricter radio reference. Their radio screening stability therefore coexists with sensitivity to beacon policy and carrier-evidence requirements.

\FloatBarrier
\subsection{Asset-level audit of the three decommissioning referrals}\label{sec:results:cases}

The three records are MDSHA-owned urban assets on C2: device 4413 on the I-695 inner loop before Providence Road, and devices 5509 and 5510 on northbound and southbound I-97 south of Hawkins Road. All three have an explicit $0\times0$ matrix and an active beacon in the inventory. The baseline tree codes that matrix as Static and routes urban Static records to decommissioning review after the ownership and rural-beacon branches. The referral therefore depends on the inventory coding and the rural restriction on beacon retention, as well as on radio support. Field verification of the sign and beacon function is required before interpreting a Static record as a removable asset.

\begin{table}[pos=!htbp]
\centering\small
\caption{Radio evidence for the three baseline referrals within 0.5 mi. T, A, and V denote T-Mobile, AT\&T, and Verizon. $J(G)$ is the percentage of observations jointly meeting the stricter reference; a dash denotes no retained observations.}
\label{tab:case_audit}
\begin{tabular}{@{}lccc@{}}
\toprule
Evidence & 4413 (ID 45) & 5509 (ID 61) & 5510 (ID 62) \\
\midrule
Observations, T/A/V & 57/0/57 & 56/0/56 & 53/0/53 \\
Spatial cells, T/A/V & 10/0/10 & 11/0/11 & 10/0/10 \\
Baseline supporting carriers & T/V & T/V & T/V \\
Stricter supporting carriers & V & T/V & T \\
$J(G)$, T/A/V (\%) & 61.4/---/100.0 & 91.1/---/89.3 & 90.6/---/77.4 \\
All-carrier variant & Insufficient & Insufficient & Insufficient \\
All-beacon variant & Retain & Retain & Retain \\
\bottomrule
\end{tabular}
\end{table}

Device 4413 has 57 complete observations for each of T-Mobile and Verizon, spanning ten grid cells per carrier. Verizon supplies the stricter same-carrier pass; T-Mobile does not. Although all retained rows meet the baseline references, only 61.4\% of T-Mobile rows meet all three stricter references together. Its footprint also contains 57 raw AT\&T records, all missing RSRQ, so none enters the complete-record screen. The first follow-up task is to inspect the original logs for recoverable fields, then collect additional measurements if needed and verify the actual approach served by the inner-loop sign.

Device 5509 has 56 complete observations and 11 cells per retained carrier. Both T-Mobile and Verizon pass the stricter marginal test, while their joint exceedance shares are 91.1\% and 89.3\%. The latter value illustrates why separate marginal $P_{10}$ tests cannot be interpreted as a 90\% joint guarantee. All 56 raw AT\&T records in the footprint lack SNR and are excluded from the screen. The northbound inventory designation identifies the intended travel direction, but the circular footprint does not establish direction-specific approach coverage.

Device 5510 has 53 complete observations and ten cells per retained carrier. T-Mobile provides the stricter marginal pass; Verizon's joint stricter-reference share is 77.4\%. Its 53 raw AT\&T records also lack SNR. Its footprint shares 78 retained source observations with device 5509, out of 140 distinct observations across the two footprints. These adjacent signs therefore do not provide two independent replications of the radio screen. Their northbound and southbound functions require separate message and approach reviews despite the shared historical radio evidence.

For all three assets, the audited inputs do not establish the messages historically served, the beacon's operational purpose, enrollment or reach of a replacement channel, delivery before the final useful decision point, an independent backup path, or comparative lifecycle costs. Those fields remain open in the decision records. The evidence supports referral for service and inventory review, with no removal authorization. A completed comparison should document feasible retention, conversion, and removal alternatives on a common cost horizon after service requirements are satisfied. No cost ranking or expert agreement is inferred from the current records.

\FloatBarrier
\section{Discussion and policy implications}\label{sec:discussion}

The Maryland application shows how evidence requirements and institutional assumptions jointly shape the screening of roadside traveler-information assets. Local connectivity measurements determine which assets have sufficient radio evidence to enter the screening process, while inventory rules influence their subsequent review categories. The practical value of the framework is its ability to make these dependencies explicit and identify the additional information needed for an agency decision. The results therefore inform both asset-review priorities and the design of procedures for managing the transition to mobile traveler information.

\subsection{Policy implications for asset retirement}

The rule audit shows how ownership and beacon policies shape the screening portfolio. Removing ownership-based retention changes 38 asset labels and increases decommissioning referrals from three to 21, whereas extending beacon retention to urban assets changes 39 labels and reduces referrals to zero. All three baseline referrals remain unchanged across the tested radio-reference and footprint variants and move to \textit{Retain} under the all-beacon alternative. Agencies can use these transition records to document the operational basis of each rule before adopting the decision tree.

All 105 DMS with complete local radio observations pass the baseline connectivity screen, after which inventory rules determine their review categories. Conditional and unmeasured assets identify priorities for follow-up work, including targeted field measurements, recovery of missing radio fields, and explicit recording of service status.

Corridor and asset-footprint assessments provide complementary information on the communication function. Corridor measurements describe the travel path, while footprints characterize conditions near individual assets. For C1, the higher all-carrier low-score share identifies approach coverage and independent communication paths as review priorities. Direction of travel, warning lead time, and available diversion points can guide the selection of evaluation footprints and subsequent service-equivalence review.

Carrier-specific results complement the best-of-three infrastructure summary by describing measured conditions on individual networks. Agencies can assess the reach of a proposed replacement through the carriers and notification services used by the intended audience. Audience review can address device access, application enrollment, alert settings, and the needs of visitors unfamiliar with local services.

\subsection{Transferability and implementation}

The transferable component of the framework is the separation of measured evidence, screening assumptions, and agency decisions. Another agency can apply this structure using a geocoded inventory, carrier-specific field observations, documented data gaps, and explicit rules for assigning review categories. The decision record should preserve the information supporting each category, the sensitivity of that category to alternative assumptions, and the rationale for the eventual agency action. This documentation allows subsequent reviewers to trace how new evidence or a change in policy affects an earlier assessment.

The framework accommodates local calibration of connectivity-score settings, carrier matching, spatial aggregation, asset footprints, and inventory branches. Agencies can select these settings to reflect communication functions, roadway geometry, network conditions, and institutional responsibilities. Evaluation footprints should represent the approaches and decision points served by each asset.

Implementation can proceed through staged review. Assets with screening results that remain stable under the tested variants can receive detailed operational assessment, while conditional and unmeasured assets can be directed toward targeted evidence collection. Once feasible alternatives have been identified, agencies can compare their lifecycle costs on a common basis. Where operationally appropriate, reversible trials could provide additional evidence while preserving necessary backup communication. The decision record can then be updated as field measurements, asset conditions, or service requirements change.

\subsection{Future validation}

Future validation should combine concurrent receiver-side radio measurements with complete message-attempt logs, including successful and failed attempts and delivery delays. Independent campaigns across carriers and seasons can evaluate delivery performance while extending spatial coverage and distinguishing no-service conditions from unavailable fields. Agency assessments can examine agreement with screening categories and compare service-feasible alternatives on a common lifecycle-cost basis.

\FloatBarrier
\section{Conclusions}\label{sec:conclusions}

This paper presents an auditable screening procedure for reviewing rural HAR and DMS assets when mobile delivery is being considered as part of the traveler-information system. The procedure combines carrier-specific drive tests from 2025, documented message receipts from 2026, and lower-tail radio measurements near each asset. Historical radio measurements spatially associated with the message-test locations provide common references for asset screening. The decision record connects local evidence status with inventory-based review categories and subsequent service-equivalence assessment.

The Maryland results show why corridor and asset evidence should be reported separately. In the primary complete-case analysis, among 3{,}723 route-specific cells constructed from tri-carrier matches, at least one carrier met the selected connectivity-score criterion in 92.8\% of cells, while all three fell below it in 7.2\%. The best-of-three share fell to 75.9\% when the score threshold increased from 30 to 50, and C1 consistently showed the greatest concentration of all-carrier low-score cells. A sensitivity analysis including incomplete records whose available RSRP or SNR values determine a zero score increased the matched set to 4{,}113 cells and yielded corresponding shares of 91.5\% and 8.5\%. Carrier-specific results accompany these best-of-three summaries.

At the asset scale, seven HAR sites had complete-case measurements within the 3-mi baseline footprint and all passed the inclusive same-carrier screen using the historically associated radio reference. Five passed both the stricter historical-reference and expanded 5-mi footprint tests, while HAR-2 and HAR-9 were sensitive to the selected test. The proposed DMS review tree assigned 58 signs to Retain, 40 to Convert, three to decommissioning review, and 18 to Inventory review. The remaining 115 signs lacked retained local complete-case drive-test evidence, including 15 corridor-assigned gaps and 100 off-corridor signs.

The decision audit shows how evidence requirements and inventory rules change the portfolio. An inventory-only scenario assigns action-review categories to 115 assets without retained local complete-case measurements, while an evidence-gated scenario places 15 conditional baseline assignments on hold. Removing the MDTA retention branch increases decommissioning referrals from three to 21; extending beacon retention to all measured, baseline-passing signs reduces them to zero. The three baseline referrals remain stable under the tested historical-reference and footprint variants. These comparisons identify the evidence requirements and institutional assumptions that drive changes in portfolio-level review categories.

The transferable contribution is a decision record that connects historical connectivity, sensitivity, inventory-based review categories, and agency judgment. Agencies can update score settings, historical radio references, asset footprints, and decision branches as new evidence becomes available while retaining a traceable screening structure. The framework prioritizes asset-specific service review and organizes the supporting evidence on message function, audience reach, delivery timing, backup provision, and lifecycle costs.

\section*{CRediT authorship contribution statement}

\textbf{Dianwei Chen}: Conceptualization, Methodology, Software, Formal analysis, Data curation, Visualization, Writing -- original draft. \textbf{Yuan-Zheng Lei}: Methodology, Writing -- review \& editing. \textbf{Zheyu Li}: Investigation, Data curation, Writing -- review \& editing. \textbf{Xianfeng Terry Yang}: Conceptualization, Methodology, Supervision, Project administration, Funding acquisition, Writing -- review \& editing.

\section*{Declaration of competing interest}

The authors declare that they have no known competing financial interests or personal relationships that could have appeared to influence the work reported in this paper.

\section*{Acknowledgments}

This research is supported by the Maryland State Highway Administration. The authors thank MDSHA CHART operations staff for access to the roadside ITS inventory, field campaigns, and review. The views expressed are those of the authors and not necessarily those of the sponsor.

\section*{Data availability}

The drive-test dataset, delivery logs, and MDSHA mobile-coverage map are agency property and are not publicly redistributable under the project data-use agreement. Aggregated statistics and screening outputs are available from the corresponding author on reasonable request, subject to agency review.

\section*{Declaration of generative AI and AI-assisted technologies in the manuscript preparation process}

During the preparation and revision of this manuscript, the authors used OpenAI Codex to assist with prose drafting and revision, development and refactoring of analysis code, data consistency checks, interpretation of analytical results, and figure-caption editing. The authors reviewed and checked the AI-assisted text, code, calculations, and interpretations, revised them as needed, and take full responsibility for the accuracy, integrity, and final content of the manuscript.

\clearpage
\bibliographystyle{cas-model2-names}
\renewcommand{\bibfont}{\fontsize{9pt}{10.5pt}\selectfont}

\bibliography{references}

\begin{thebibliography}{22}
\expandafter\ifx\csname natexlab\endcsname\relax\def\natexlab#1{#1}\fi
\providecommand{\url}[1]{\texttt{#1}}
\providecommand{\href}[2]{#2}
\providecommand{\path}[1]{#1}
\providecommand{\DOIprefix}{doi:}
\providecommand{\ArXivprefix}{arXiv:}
\providecommand{\URLprefix}{URL: }
\providecommand{\Pubmedprefix}{pmid:}
\providecommand{\doi}[1]{\href{http://dx.doi.org/#1}{\path{#1}}}
\providecommand{\Pubmed}[1]{\href{pmid:#1}{\path{#1}}}
\providecommand{\bibinfo}[2]{#2}
\ifx\xfnm\relax \def\xfnm[#1]{\unskip,\space#1}\fi
\bibitem[{Adarsh et~al.(2021)Adarsh, Nekrasov, Paul, Mangla, Gupta,
  Vigil-Hayes, Zegura and Belding}]{adarsh2021coverage}
\bibinfo{author}{Adarsh, V.}, \bibinfo{author}{Nekrasov, M.},
  \bibinfo{author}{Paul, U.}, \bibinfo{author}{Mangla, T.},
  \bibinfo{author}{Gupta, A.}, \bibinfo{author}{Vigil-Hayes, M.},
  \bibinfo{author}{Zegura, E.}, \bibinfo{author}{Belding, E.},
  \bibinfo{year}{2021}.
\newblock \bibinfo{title}{Coverage is not binary: Quantifying mobile broadband
  quality in urban, rural, and tribal contexts}, in: \bibinfo{booktitle}{2021
  International Conference on Computer Communications and Networks (ICCCN)},
  \bibinfo{organization}{IEEE}. pp. \bibinfo{pages}{1--9}.
\bibitem[{{American Association of State Highway and Transportation Officials
  (AASHTO)}(2011)}]{american2011transportation}
\bibinfo{author}{{American Association of State Highway and Transportation
  Officials (AASHTO)}}, \bibinfo{year}{2011}.
\newblock \bibinfo{title}{Transportation asset management guide: a focus on
  implementation}.
\newblock \bibinfo{journal}{Washington, DC, 48pp} .
\bibitem[{Amin-Naseri et~al.(2018)Amin-Naseri, Chakraborty, Sharma, Gilbert and
  Hong}]{amin2018evaluating}
\bibinfo{author}{Amin-Naseri, M.}, \bibinfo{author}{Chakraborty, P.},
  \bibinfo{author}{Sharma, A.}, \bibinfo{author}{Gilbert, S.B.},
  \bibinfo{author}{Hong, M.}, \bibinfo{year}{2018}.
\newblock \bibinfo{title}{Evaluating the reliability, coverage, and added value
  of crowdsourced traffic incident reports from waze}.
\newblock \bibinfo{journal}{Transportation research record}
  \bibinfo{volume}{2672}, \bibinfo{pages}{34--43}.
\bibitem[{Bean et~al.(2015)Bean, Sutton, Liu, Madden, Wood and
  Mileti}]{bean2015study}
\bibinfo{author}{Bean, H.}, \bibinfo{author}{Sutton, J.}, \bibinfo{author}{Liu,
  B.F.}, \bibinfo{author}{Madden, S.}, \bibinfo{author}{Wood, M.M.},
  \bibinfo{author}{Mileti, D.S.}, \bibinfo{year}{2015}.
\newblock \bibinfo{title}{The study of mobile public warning messages: A
  research review and agenda}.
\newblock \bibinfo{journal}{Review of Communication} \bibinfo{volume}{15},
  \bibinfo{pages}{60--80}.
\bibitem[{Broniewicz and Ogrodnik(2025)}]{broniewicz2025application}
\bibinfo{author}{Broniewicz, E.}, \bibinfo{author}{Ogrodnik, K.},
  \bibinfo{year}{2025}.
\newblock \bibinfo{title}{Application potential of {MCDM/MCDA} methods in
  transport---literature review and case study}.
\newblock \bibinfo{journal}{Sustainability} \bibinfo{volume}{17},
  \bibinfo{pages}{7671}.
\bibitem[{Cain et~al.(2024)Cain, Sutton and Olson}]{cain2024wireless}
\bibinfo{author}{Cain, L.B.}, \bibinfo{author}{Sutton, J.},
  \bibinfo{author}{Olson, M.K.}, \bibinfo{year}{2024}.
\newblock \bibinfo{title}{Wireless emergency alerts and organisational
  response: Instructing and adjusting information in alerts}.
\newblock \bibinfo{journal}{Journal of Contingencies and Crisis Management}
  \bibinfo{volume}{32}, \bibinfo{pages}{e12516}.
\bibitem[{Chang and Rochon(2009)}]{chang2009performance}
\bibinfo{author}{Chang, G.L.}, \bibinfo{author}{Rochon, S.},
  \bibinfo{year}{2009}.
\newblock \bibinfo{title}{Performance evaluation and benefit analysis for
  {CHART}}.
\newblock \bibinfo{journal}{Coordinated Highways Action Response Team, Maryland
  Department of Transportation} .
\bibitem[{Chatterjee et~al.(2002)Chatterjee, Hounsell, Firmin and
  Bonsall}]{chatterjee2002driver}
\bibinfo{author}{Chatterjee, K.}, \bibinfo{author}{Hounsell, N.},
  \bibinfo{author}{Firmin, P.}, \bibinfo{author}{Bonsall, P.},
  \bibinfo{year}{2002}.
\newblock \bibinfo{title}{Driver response to variable message sign information
  in london}.
\newblock \bibinfo{journal}{Transportation research part C: Emerging
  technologies} \bibinfo{volume}{10}, \bibinfo{pages}{149--169}.
\bibitem[{Dudek(2004)}]{dudek2004changeable}
\bibinfo{author}{Dudek, C.L.}, \bibinfo{year}{2004}.
\newblock \bibinfo{title}{Changeable message sign operation and messaging
  handbook} .
\bibitem[{Edwards et~al.(2022)Edwards, Liu, Hughes, Gao, Goodrich, Barner and
  Herrera}]{edwards2022best}
\bibinfo{author}{Edwards, F.}, \bibinfo{author}{Liu, K.},
  \bibinfo{author}{Hughes, A.L.}, \bibinfo{author}{Gao, J.Z.},
  \bibinfo{author}{Goodrich, D.}, \bibinfo{author}{Barner, A.},
  \bibinfo{author}{Herrera, R.}, \bibinfo{year}{2022}.
\newblock \bibinfo{title}{Best practices in disaster public communications:
  Evacuation alerting and social media} .
\bibitem[{Kondyli et~al.(2026)Kondyli, Schrock, Mahajan, Masud, Karki, Shrestha
  and Tabatabaei}]{kondyli2026creating}
\bibinfo{author}{Kondyli, A.}, \bibinfo{author}{Schrock, S.D.},
  \bibinfo{author}{Mahajan, K.}, \bibinfo{author}{Masud, S.S.B.},
  \bibinfo{author}{Karki, A.M.}, \bibinfo{author}{Shrestha, S.},
  \bibinfo{author}{Tabatabaei, M.}, \bibinfo{year}{2026}.
\newblock \bibinfo{title}{Creating effective graphic changeable dynamic message
  sign ({DMS}) messaging} .
\bibitem[{Leveson(2012)}]{leveson2012engineering}
\bibinfo{author}{Leveson, N.}, \bibinfo{year}{2012}.
\newblock \bibinfo{title}{Engineering a safer world: applying systems thinking
  to safety}.
\newblock \bibinfo{publisher}{MIT Press}.
\bibitem[{Mangla et~al.(2021)Mangla, Showalter, Adarsh, Jones, Vigil-Hayes,
  Belding and Zegura}]{mangla2021tale}
\bibinfo{author}{Mangla, T.}, \bibinfo{author}{Showalter, E.},
  \bibinfo{author}{Adarsh, V.}, \bibinfo{author}{Jones, K.},
  \bibinfo{author}{Vigil-Hayes, M.}, \bibinfo{author}{Belding, E.},
  \bibinfo{author}{Zegura, E.}, \bibinfo{year}{2021}.
\newblock \bibinfo{title}{A tale of three datasets: Towards characterizing
  mobile broadband access in the {United States}}.
\newblock \bibinfo{journal}{arXiv preprint arXiv:2102.07288} .
\bibitem[{McKay and Senesi(2022)}]{mckay2022applying}
\bibinfo{author}{McKay, G.}, \bibinfo{author}{Senesi, C.},
  \bibinfo{year}{2022}.
\newblock \bibinfo{title}{Applying transportation asset management to
  intelligent transportation systems assets: A primer} .
\bibitem[{Mossberger et~al.(2003)Mossberger, Tolbert and
  Stansbury}]{mossberger2003virtual}
\bibinfo{author}{Mossberger, K.}, \bibinfo{author}{Tolbert, C.J.},
  \bibinfo{author}{Stansbury, M.}, \bibinfo{year}{2003}.
\newblock \bibinfo{title}{Virtual inequality: Beyond the digital divide}.
\newblock \bibinfo{publisher}{Georgetown University Press}.
\bibitem[{Olson et~al.(2024)Olson, Sutton, Cain and Waugh}]{olson2024decade}
\bibinfo{author}{Olson, M.K.}, \bibinfo{author}{Sutton, J.},
  \bibinfo{author}{Cain, L.B.}, \bibinfo{author}{Waugh, N.},
  \bibinfo{year}{2024}.
\newblock \bibinfo{title}{A decade of wireless emergency alerts: A longitudinal
  assessment of message content and completeness}.
\newblock \bibinfo{journal}{Journal of Contingencies and Crisis Management}
  \bibinfo{volume}{32}, \bibinfo{pages}{e12518}.
\bibitem[{Perrin(2021)}]{perrin2021mobile}
\bibinfo{author}{Perrin, A.}, \bibinfo{year}{2021}.
\newblock \bibinfo{title}{Mobile technology and home broadband 2021}.
\newblock \bibinfo{journal}{Pew Research Center} .
\bibitem[{Raman et~al.(2006)Raman, Zimmerman, Kell and
  Bausher}]{raman2006communicating}
\bibinfo{author}{Raman, M.}, \bibinfo{author}{Zimmerman, C.A.},
  \bibinfo{author}{Kell, T.}, \bibinfo{author}{Bausher, C.},
  \bibinfo{year}{2006}.
\newblock \bibinfo{title}{Communicating with the public using {ATIS} during
  disasters-concept of operations} .
\bibitem[{Sui and Young(2014)}]{sui2014impact}
\bibinfo{author}{Sui, Y.}, \bibinfo{author}{Young, R.}, \bibinfo{year}{2014}.
\newblock \bibinfo{title}{Impact of dynamic message signs on speeds observed on
  a rural interstate}.
\newblock \bibinfo{journal}{Journal of Transportation Engineering}
  \bibinfo{volume}{140}, \bibinfo{pages}{04014020}.
\bibitem[{Van~Dijk(2020)}]{van2020digital}
\bibinfo{author}{Van~Dijk, J.A.G.M.}, \bibinfo{year}{2020}.
\newblock \bibinfo{title}{The digital divide}.
\newblock \bibinfo{publisher}{Polity}.
\bibitem[{Wardman et~al.(1997)Wardman, Bonsall and Shires}]{wardman1997driver}
\bibinfo{author}{Wardman, M.}, \bibinfo{author}{Bonsall, P.},
  \bibinfo{author}{Shires, J.}, \bibinfo{year}{1997}.
\newblock \bibinfo{title}{Driver response to variable message signs: a stated
  preference investigation}.
\newblock \bibinfo{journal}{Transportation Research Part C: Emerging
  Technologies} \bibinfo{volume}{5}, \bibinfo{pages}{389--405}.
\bibitem[{Weaver et~al.(2024)Weaver, Ahmed, Jannat, Olko and
  Arnold}]{weaver2024systematic}
\bibinfo{author}{Weaver, S.}, \bibinfo{author}{Ahmed, A.},
  \bibinfo{author}{Jannat, M.}, \bibinfo{author}{Olko, S.},
  \bibinfo{author}{Arnold, M.}, \bibinfo{year}{2024}.
\newblock \bibinfo{title}{A systematic approach on {CMS} messaging selection
  during nonrecurring events: Decision tree} .

\end{thebibliography}

\end{document}